\documentclass[intlimits,twoside,a4paper]{article}

\usepackage{amsmath,amssymb}
\usepackage{graphicx}

\usepackage[T2A]{fontenc}
\usepackage[cp1251]{inputenc}

\usepackage[eqsecnum]{cmpj2}

\issue{2016}{19}{4}{43703}
\doinumber{10.5488/CMP.19.43703}
\title[The effects of strong correlations on the band structure of Ag$_{8}$SnSe$_{6}$ argyrodite]%
{The effects of strong correlations on the band structure of Ag$_{8}$SnSe$_{6}$ argyrodite%
}
\author[S.V. Syrotyuk, I.V. Semkiv, H.A. Ilchuk, V.M. Shved]{S.V. Syrotyuk,
        I.V. Semkiv,
        H.A. Ilchuk,
        V.M. Shved}
\address{ Lviv Polytechnic National University, 12 Bandera St., 79013 Lviv, Ukraine
}

\authorcopyright{S.V. Syrotyuk, I.V. Semkiv, H.A. Ilchuk, V.M. Shved, 2016}
\date{Received June 24, 2016, in final form September 27, 2016}

\begin{document}

\maketitle

\begin{abstract}

The electronic energy band spectra, as well as partial and total density of electronic states of the crystal argyrodite Ag$_{8}$SnSe$_{6}$ have been evaluated within the projector augmented waves (PAW) approach by means of the ABINIT code. The one-electron energies have been evaluated using two functionals for exchange-correlation energy. The first one is the generalized gradient approximation (GGA) approach. The second one is the hybrid functional PBE0 composed of the semilocal GGA part and Hartree-Fock exact exchange non-local energy for strongly correlated $4d$ electrons of Ag atom.  The second approach eliminates the Coulomb self-interaction of the Ag $4d$ electrons. This leads to a significant restructuring of the energy bands in the filled valence part and to an improved location of the Ag $4d$-states on the energy scale, and the resulting value of the band gap is well compared with experiment. The effects of strong correlation on the electronic structure of the crystal argyrodite are considered here for the first time.

\pacs 71.15.Ap, 71.15.Mb, 71.20.Nr, 71.27.+a
\end{abstract}

\section{Introduction}

Ternary chalcogenide compounds have attracted much research interest due to their possible applications in photovoltaic and optoelectronic devices, as well as in non-volatile memories, especially in the resistive random access memory (RRAM) with low energy consumption and high speed of action \cite{Lee11,Waser09}. The phase-change chalcogenides have been used as resistive memory cell materials \cite{Kozicki05}. Compounds represented by the formulae A$^{\text I}_{8}$B$^{\text{IV}}$X$^{\text{II}}_{6}$ (A$^{\text I}$ = Cu, Ag; B$^{\text{IV}}$ = Si, Ge, Sn; X$^{\text{II}}$ = S, Se, Te) \cite{Kuhs79} belong to the argyrodite family.

The argyrodite Ag$_{8}$SnSe$_{6}$ is a narrow band gap semiconductor with energy band gap of \(0.82\)~eV \cite{Bendorius75,Gorochov68}. The absorption spectrum of the system Ag$_{8}$SnSe$_{6}$ is characterized by a sharp absorption peak in the IR region. The structure of Ag$_{8}$SnSe$_{6}$ was described by orthorhombic Bravais lattice (space group \emph{Pmn}2$_{1}$) below the phase transition point \cite{Semkiv16,Gulay02}. The group-theoretical analysis shows that Ag$_{8}$SnSe$_{6}$ compound contains 87 optical and 3 acoustic oscillating modes \cite{Semkiv15}.

The interest in the argyrodite materials is caused by their mixed electronic-ionic conduction and low-temperature phase transition $\beta'\leftrightarrow\gamma$ (at 83\textcelsius) \cite{Chekaylo11}.

The electronic structure of argyrodite Ag$_{8}$SnSe$_{6}$ has been evaluated within the plane-wave pseudopotential formalism \cite{Semkiv16}, without taking into account the effects of strong correlations.

It is clear that the $4d$ electrons of Ag are localized in a narrow energy range of the valence band. Therefore, our task will be to clarify the important points related to the localization of electrons in the valence band, depending on whether the strong correlation of $d$ electrons are taken into account or not. The local exchange-correlation potential fails in this system and can lead to higher energy band position $E_{4d}(k)$ and to negative band gap values $E_{\text g}\leqslant 0$. For example, using the exchange-correlation potential without strong correlation leads to the metal electronic structure in CdO crystal \cite{Syrotyuk131}. Strong electron correlation being taken into account provides a good agreement with experimental direct and indirect band gap values. In works \cite{Syrotyuk131,Syrotyuk132} it was also concluded that strong correlation of the $d$~electrons of transition element impurities has a considerable effect on the electronic structure of CdO and AlN crystals.

The effects of strongly correlated electrons in a crystal Ag$_{8}$SnSe$_{6}$ have not been considered earlier. So the main purpose of this study is to elucidate the following issues: 1) what is the impact of a strong correlation on the value of the band gap? 2) what are the changes in the electronic structure caused by strong correlations?

We hope to get the answers to these questions having performed the calculations of the electronic structure taking into account strong correlation of the Ag $4d$ electrons.

\section{Theory}

The projector augmented waves (PAW) method \cite{Blochl94} combines features of pseudopotential approach and augmented plane waves method. In PAW method, the crystal is divided into two parts: 1) region inside the spheres built around the atoms --– an augmentation region $(\Omega_{\text{aug}})$; 2) region outside the spheres $(\Omega_{\text{int}})$ --- an interstitial region. The wave functions are represented differently in these regions of the crystal.

For the core electron shells, the wave function is expanded into atomic partial waves $|\psi_{\text{aug}}\rangle=\sum\nolimits_{i}c_{i}\varphi_{i}$. Inside the sphere $(\Omega_{\text{aug}})$, but in its distant part, the pseudo-wave function is expanded by auxiliary partial waves $|\tilde{\psi}_{\text{aug}}\rangle=\sum\nolimits_{i}c_{i}\tilde{\varphi}_{i}$. Between the spheres in the crystal smooth pseudo-wave function is well approximated in terms of plane waves basis ($\thicksim10^{3}$) $|\tilde{\psi}\rangle=\sum\nolimits_{G}a_{G}(k)|k+G\rangle$.

In the crystal, all the electron wave function of an electron consists of three components described above as follows:
\begin{equation}
\label{2.1}
|\psi\rangle = |\tilde{\psi}\rangle+\sum\limits_{i}(|\varphi_{i}\rangle-|\tilde{\varphi}_{i}\rangle)\langle\tilde{p}_{i}|\tilde{\psi}\rangle,
\end{equation}
where $\langle\tilde{p}_{i}|$ is the projector function.

Transformation from pseudo-wave to all electron function is determined by operator $\tau$,
\begin{equation}
\label{2.2}
|{\psi}_{n}\rangle=\tau|\tilde{\psi}_{n}\rangle,
\end{equation}
then, stationary Schr\"{o}dinger equation takes the following form:
\begin{equation}
\label{2.3}
\tau^{\ast}H\tau|\tilde{\psi}_{n}\rangle=\tau^{\ast}\tau|\tilde{\psi}_{n}\rangle\varepsilon_{n}\,.
\end{equation}
Exchange-correlation hybrid functional PBE0 \cite{Ernzerhof99} is obtained from the equation
\begin{equation}
\label{2.4}
E^{\text{PBE}0}_{\text{xc}}[n]=E^{\text{PBE}}_{\text{xc}}[n]+\alpha\left(E^{\text{HF}}_{\text{x}}[\psi_{\text{sel}}]-E^{\text{PBE}}_{\text{x}}[n_{\text{sel}}]\right),
\end{equation}
where $E^{\text{PBE}}_{\text{xc}}[n]$ is the exchange-correlation functional in the generalized gradient approximation \cite{Perdew96}, \linebreak $E^{\text{HF}}_{\text{x}}[\psi_{\text{sel}}]$ is the exchange energy in Hartree-Fock theory, $\psi_{\text{sel}}$ and $n_{\text{sel}}$ is the wave function and electron density of selected electrons \cite{Tran06}, i.e., \(4d\) electrons of Ag.

\section{Calculation details}

ABINIT program \cite{Gonze09} was used for calculation of electronic properties of argyrodite crystal. The structure of $\beta$ modification of the crystal was obtained from experimental study \cite{Gulay02}. The argyrodite crystalizes in orthorhombic crystalline form with \emph{Pmn}2$_{1}$ space group. The unit cell parameters are defined as \(a=0.79168(6)\)~nm, \(b=0.78219(6)\)~nm and \(c=1.10453(8)\)~nm, respectively. The unit cell of the crystal consists of two formula units (\(Z=2\)) containing 30 atoms.

The projector augmented waves method (PAW) was used for solving the system of equations (\ref{2.3}) with respect to crystal eigenenergies $\varepsilon_{n}$ and pseudo-wave functions $\tilde{\psi}_{n}$. The PAW basis states have been generated by means of the AtomPAW program \cite{Holzwarth01} for the following configurations of the valence basis states: \(4s^{2}5s^{1}4p^{6}4d^{10}\) for Ag, \(4s^{2}5s^{2}4p^{6}5p^{2}4d^{10}\) for Sn and \(3s^{2}4s^{2}3p^{6}4p^{4}\) for Se.
The system of equations was solved self-consistently and the full convergence of the binding energy of the unit cell was achieved with the accuracy of \(10^{-7}\)~Ha.

\section{Results and discussion}

The calculations were performed with three values of \(\alpha\) parameter such as $0.0$, $0.25$ and $0.33$. The authors \cite{Ernzerhof99, Tran06} have recommended the value of \(\alpha=0.25\).

\begin{figure}[!b]
\vspace{-2mm}
\begin{minipage}[h]{0.49\linewidth}
\center{\includegraphics[width=1\linewidth]{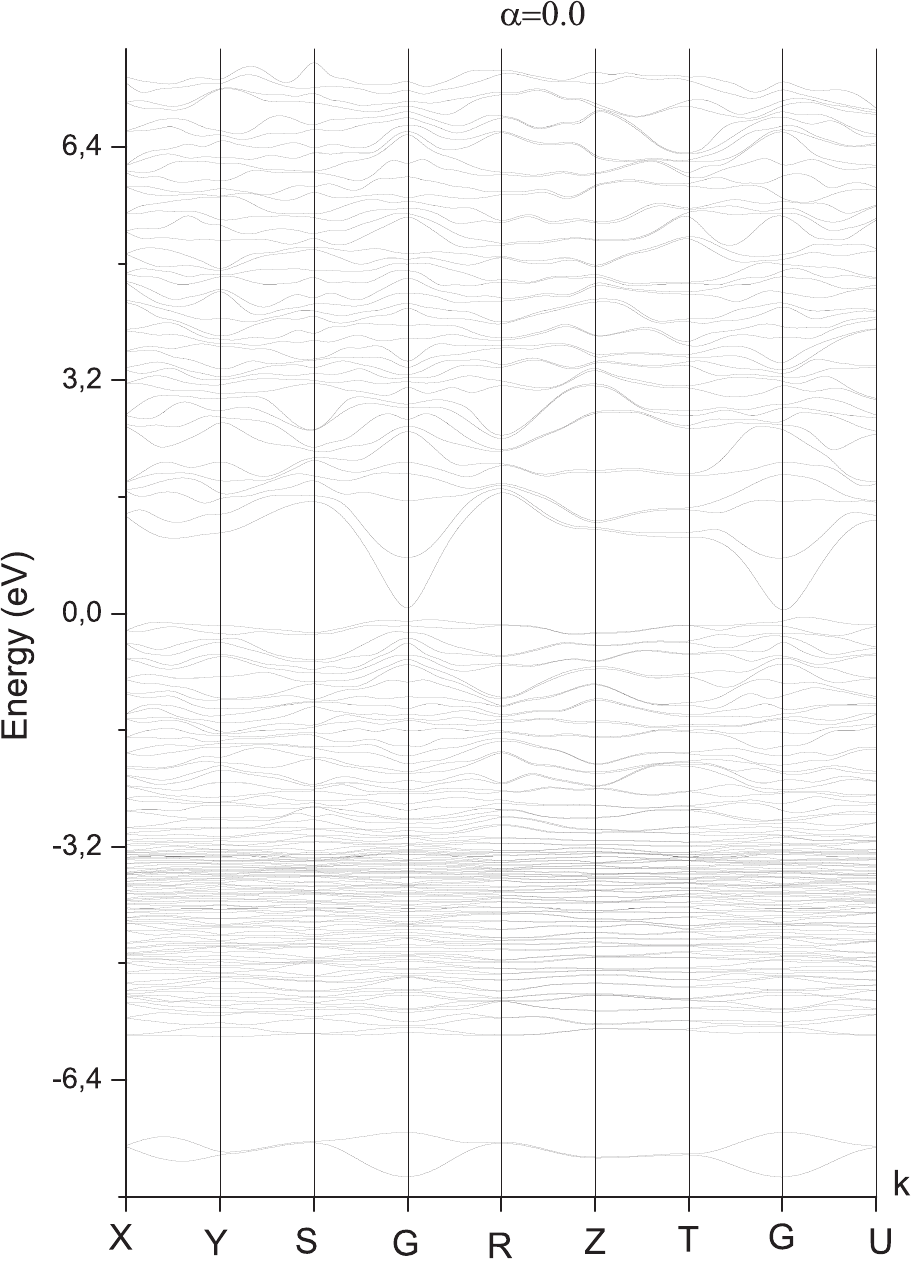} \\(a)}
\end{minipage}
\hfill
\begin{minipage}[h]{0.49\linewidth}
\center{\includegraphics[width=1\linewidth]{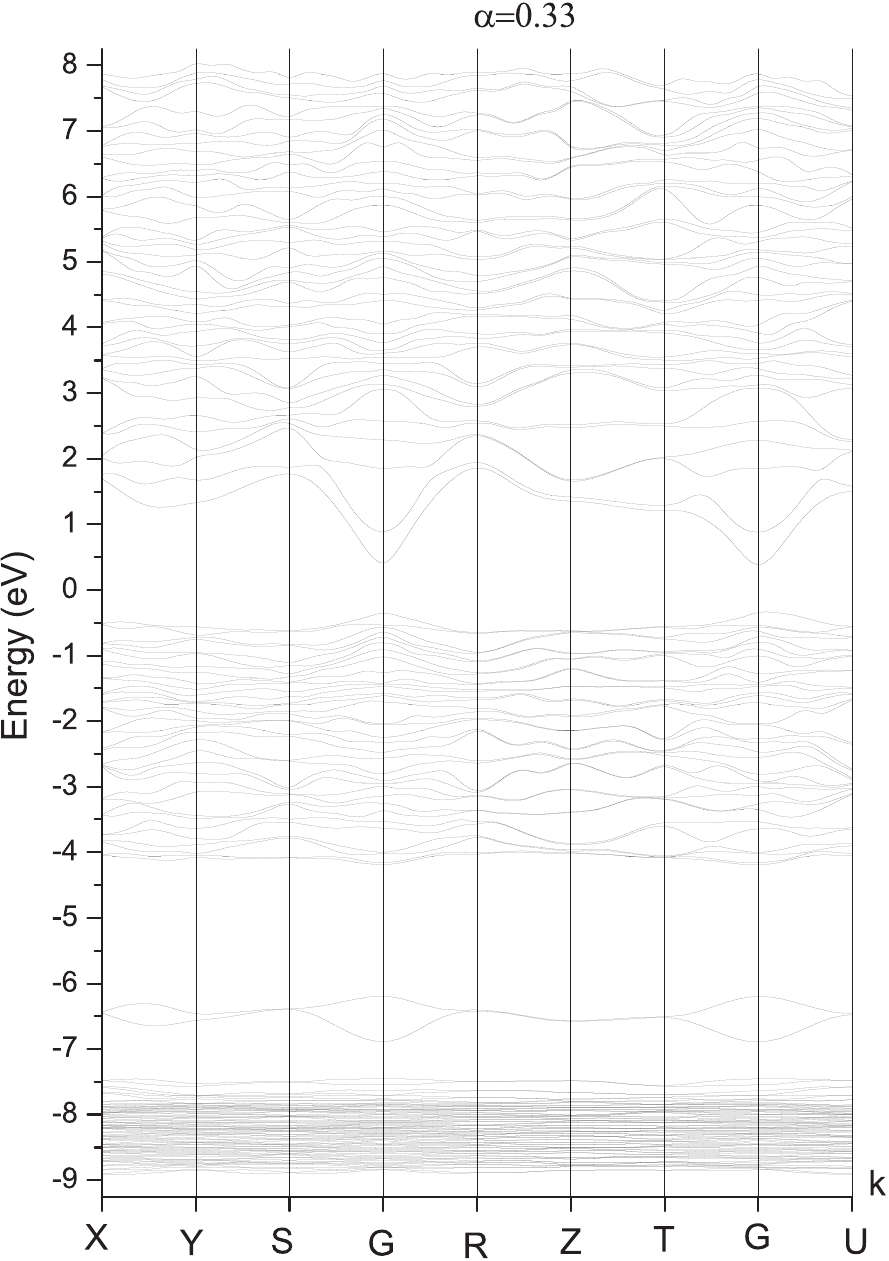} \\(b)}
\end{minipage}
\caption{The energy band structure of Ag$_{8}$SnSe$_{6}$ crystal evaluated for  $\alpha=0$ (a) and  $\alpha=0.33$ (b).}
\label{fig1}
\end{figure}

The energy band structure of Ag$_{8}$SnSe$_{6}$ is presented in figure~\ref{fig1}. The result obtained with exchange-correlation potential within GGA is shown in figure~\ref{fig1}, with the coefficient value $\alpha=0$. In this case we found the band gap value of $E_{\text g}=0.05$~eV for the Ag$_{8}$SnSe$_{6}$ crystal. The electron dispersion curves with local Coulomb correlation of Ag $4d$ electrons are presented in figure~\ref{fig1} on the right. For this calculation, the coefficient $\alpha=0.33$ was used in equation (\ref{2.4}).

Comparing two $E(\textbf{k})$ spectra (figure~\ref{fig1}) we found that the consideration of strong correlation of Ag $4d$ electrons leads to quantitative and qualitative changes of the energy band parameters. For example, the band gap value of Ag$_{8}$SnSe$_{6}$ crystal is  $E_{\text g}=0.66$~eV at $\Gamma$ point in the first Brillouin zone which is in good agreement with experimental value  $E_{\text g}=0.82$~eV \cite{Gorochov68}. For comparison, we also give a value of the bandgap $E_{\text g}=0.54$~eV, derived with the recommended option $\alpha=0.25$. According to our calculations, the considered material is characterized by a direct fundamental interband gap $\Gamma$--$\Gamma$. The top of the valence band is formed by $p$-states of Se, and at the bottom of the conduction band there dominate the $s$-states of Sn, so the dipole interband optical transition $s$--$p$ is allowed. Thus, this is also an optical gap.

The comparison between figure~\ref{fig1}~(a) and \ref{fig1}~(b) shows a significant renormalization of dispersion curves in the valence band. Below the Fermi level, the electronic states are located in the energy range $0.0\leqslant E\leqslant6.0$ as illustrated in figure~\ref{fig1}~(a). The strong correlation of Ag $4d$ electrons leads to the splitting this broad band of electronic states into two groups of energy bands which is shown in figure~\ref{fig1}~(b). The dispersion curves [figure~\ref{fig1}~(a) and \ref{fig1}~(b)] exhibit a significant difference of electronic energy band spectra obtained within GGA formalism for $\alpha=0$ in equation (\ref{2.4}) and with the strong correlation of Ag $4d$ electrons (\(\alpha=0.33\)) being taken into account.

A significant qualitative change in the electron energy spectrum, as can be seen from figure~\ref{fig1}~(a) and \ref{fig1}~(b), is obviously caused by a partial removal of the Coulomb's self-interaction from the exchange-correlation potential in equation (\ref{2.4}), setting a non-zero coefficient $\alpha$. Self-interaction leads to systematic errors in the band gap values of semiconductors and dielectrics \cite{Lebegue03,Schlipf11}.

\begin{figure}[!h]
\begin{minipage}[h]{0.49\linewidth}
\center{\includegraphics[width=1\linewidth]{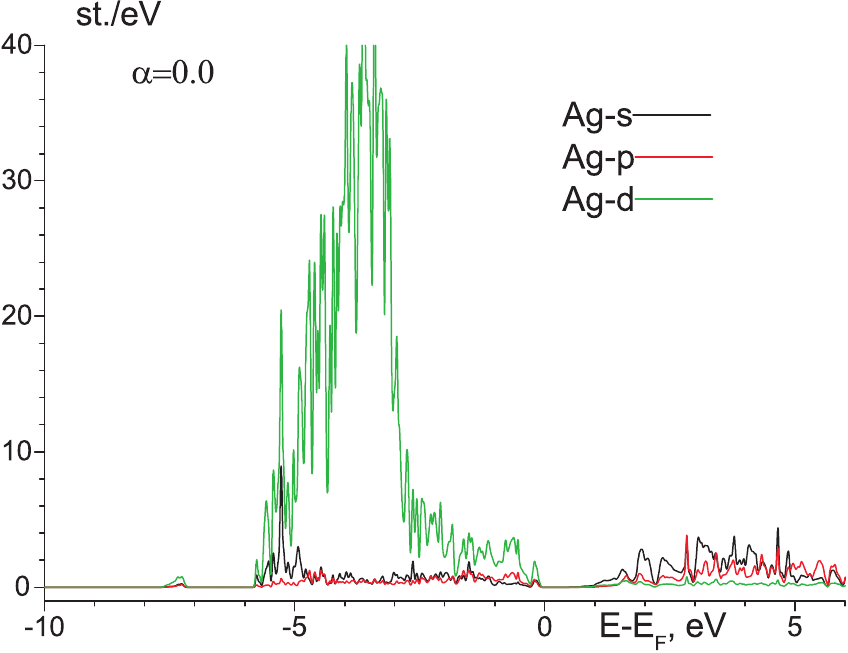} \\(a)}
\end{minipage}
\hfill
\begin{minipage}[h]{0.49\linewidth}
\center{\includegraphics[width=1\linewidth]{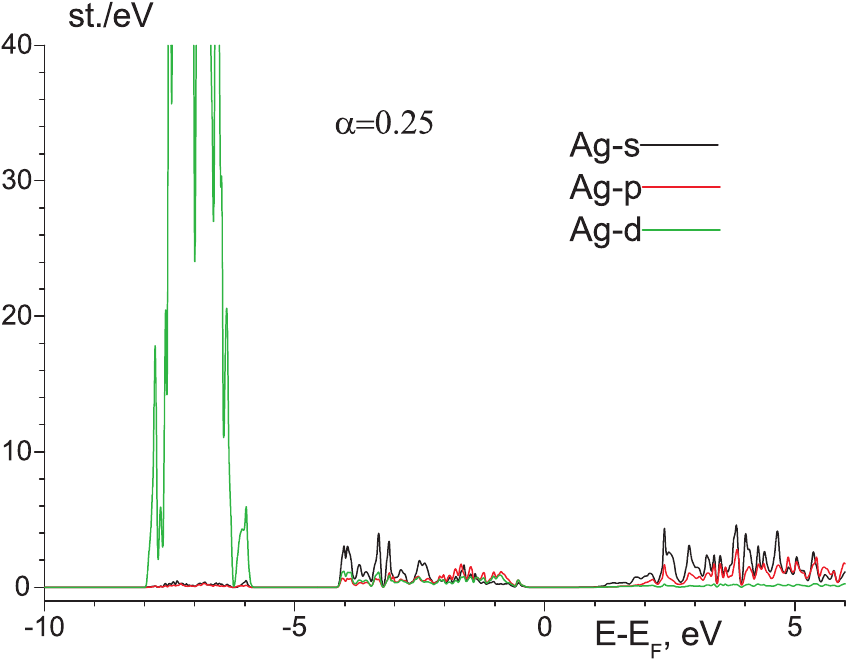} \\(b)}
\end{minipage}
\caption{(Color online) The partial density of states of Ag evaluated (a) within GGA ($\alpha=0$) and (b) using the strong correlation of Ag $4d$ electrons ($\alpha=0.25$) for Ag$_{8}$SnSe$_{6}$ crystal.}
\label{fig2}
\end{figure}

Figure~\ref{fig2} shows that Ag $4d$ electrons are localized in a narrow energy range and are characterized by high values of density of states. The energy range of about $\Delta E_{4d}\approx3$~eV is obtained within GGA method. Using the exchange-correlation potential (\ref{2.4}) with a mixing coefficient of ($\alpha=0.25$) leads to a significant decrease of this range to about $\Delta E_{4d}\approx2$~eV. Moreover, the center of Ag $4d$ electron band is shifted to lower values of energy (approximately to $3.5$~eV).

\begin{figure}[!b]
\begin{minipage}[h]{0.49\linewidth}
\center{\includegraphics[width=1\linewidth]{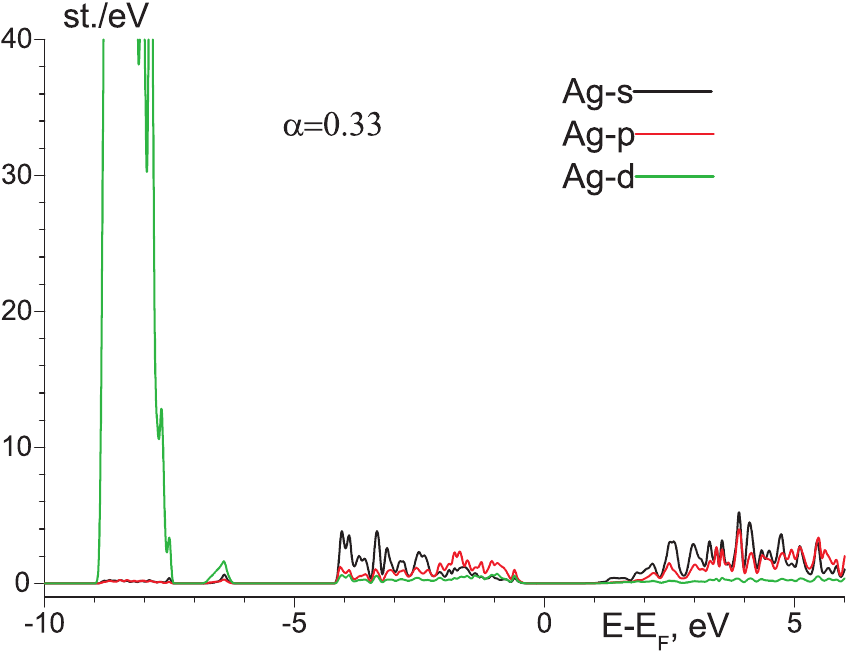} \\ (a)}
\end{minipage}
\hfill
\begin{minipage}[h]{0.49\linewidth}
\center{\includegraphics[width=1\linewidth]{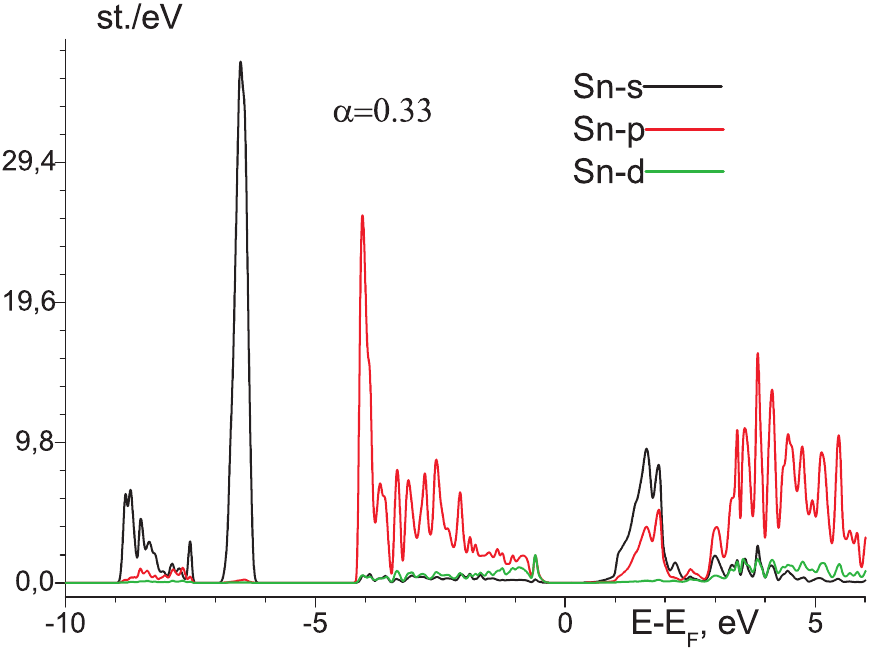} \\ (b)}
\end{minipage}
\caption{(Color online) The partial density of states of Ag (a) and Sn (b) evaluated using the strong correlation of Ag $4d$ electrons ($\alpha=0.33$) for Ag$_{8}$SnSe$_{6}$ crystal.}
\label{fig3}
\end{figure}

As shown in figure~\ref{fig3}, the partial density of states of Ag and Sn are evaluated using the mixing parameter of $\alpha=0.33$ for the exchange-correlation potential. Consequently, the densities of states are obtained with cancellation of self-interaction of Ag $4d$ electrons which are localized in narrow energy bands.

\begin{figure}[!t]
\begin{minipage}[h]{0.49\linewidth}
\center{\includegraphics[width=1\linewidth]{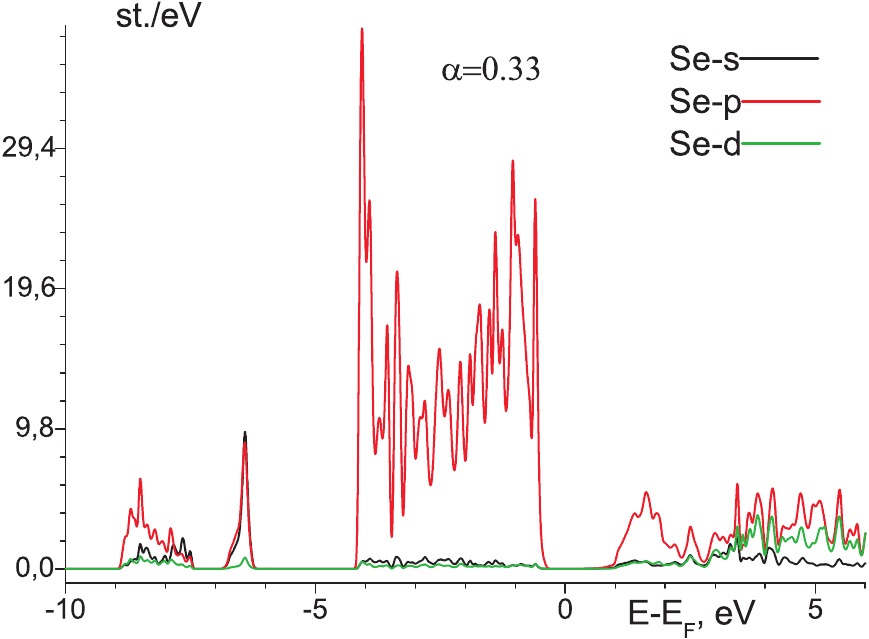} \\(a)}
\end{minipage}
\hfill
\begin{minipage}[h]{0.49\linewidth}
\center{\includegraphics[width=1\linewidth]{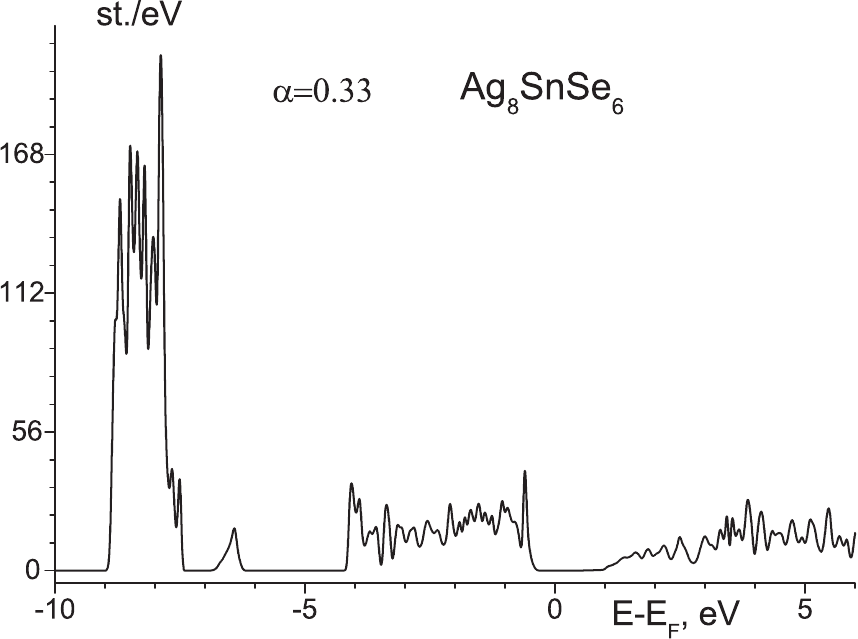} \\(b)}
\end{minipage}
\caption{(Color online) The partial and total density of states of Se (a) and Ag$_{8}$SnSe$_{6}$ crystal (b), respectively, evaluated using the strong correlation of Ag $4d$ electrons ($\alpha=0.33$).}
\label{fig4}
\end{figure}

The partial density of states presented in figure~\ref{fig3}, ~\ref{fig4} shows a significant domination of $p$-states of Se and Sn at the top of the valence band. The contribution of $s$- and $p$-states of Ag is small in this energy region. Consequently, the top of the valence band is formed from the hybridized $p$-states of Se, Sn, Ag and Ag $s$-states.

The bottom of the conduction band consists of the hybridized $p$-states of Sn and $s$-, $p$-states of Sn and~Se. The total density of states (figure~\ref{fig4}) illustrates that Ag $4d$ electrons are localized in a narrow energy bands. These Ag $4d$ electrons create a strongly correlated subsystem, which may be properly described by means of the semilocal GGA exchange-correlation potential and nonlocal Hartree-Fock exchange potential, with nonzero mixing parameter of $\alpha$ in equation (\ref{2.4}).

\section{Conclusions}

The energy band structure of Ag$_{8}$SnSe$_{6}$ crystal was obtained within the two theoretical models for exchange-correlation functional: a generalized gradient approximation GGA-PBE and a hybrid functional PBE0. In PBE0 approach, the first term describes $s$ and $p$ electrons within GGA formalism and the second one represents the Hartree-Fock energy for \(4d\) electrons of Ag. The band gap value obtained from GGA formalism is very small. Taking into account the local strong correlation of Ag \(4d\) electrons leads to the band gap value of \(0.66\)~eV which is much closer to the experiment. Involving the Hartree-Fock energy in the calculation leads to the elimination of Coulomb self-interaction and to an increase of the band gap value. Therefore, we can conclude that the wave functions of Ag $4d$ electrons, on the one hand, and the states localized at the top of the valence band, and also lying near the bottom of the conduction band, such as $p$-states of Se and $s$- and $p$-states of Sn, on the other hand, reveal a considerable hybridization.

\ukrainianpart

\title{Вплив ефекту сильних кореляцій на зонну структуру аргіродиту Ag$_{8}$SnSe$_{6}$}
\author{С.В. Сиротюк, І.В. Семків, Г.А. Ільчук, В.М. Швед}
\address{ Національний університет ``Львiвська полiтехнiка'', вул. С. Бандери, 12, 79013 Львiв, Україна
}

\makeukrtitle

\begin{abstract}
\tolerance=3000%
Електронні енергетичні спектри, а також парціальні та повна щільності електронних станів у $\beta$-фазі кристалу аргіродиту Ag$_{8}$SnSe$_{6}$ отримані в базисі проекційних приєднаних хвиль (PAW) за допомогою програми ABINIT. Одночастинкові енергії електронів були знайдені з використанням двох функціоналів обмінно-кореляційної енергії. Перший грунтується на узагальненому градієнтному наближенні (GGA). Другий є гібридним функціоналом, що складається з напівлокальної складової, сформульованої у підході GGA, та з точного нелокального обмінного потенціала Хартрі-Фока сильно скорельованих $4d$ електронів~Ag. Другий підхід усуває кулонівську самодію $4d$ електронів~Ag. Це приводить до значної перебудови структури енергетичних зон у її заповненій валентній частині, більш правильного розміщення $4d$-станів Ag за енергетичною шкалою, а отримане значення ширини забороненої зони виявляє добре зіставлення з експериментом. Вплив сильних кореляцій на електронну структуру кристала аргіродиту розглядається тут вперше.

\keywords розрахунки електронної структури, сильно скорельовані електрони, точний обмін для скорельованих електронів, проекційно приєднані хвилі PAW, халькогеніди, аргіродити

\end{abstract}

\end{document}